\documentclass[conference]{IEEEtran}

\usepackage[letterpaper, left=1.01in, right=1.01in, bottom=1in, top=0.75in]{geometry}
\usepackage{graphicx}
\usepackage{subfigure}
\usepackage{amsmath}
\usepackage{xcolor}

\newcommand{\fig}[1]{Fig.~\ref{#1}}

\begin{document}
%
\title{Aggressive Fragmentation Strategy for Enhanced Network Performance in Dense LPWANs}

\author{\IEEEauthorblockN{Ioana Suciu}
\IEEEauthorblockA{Worldsensing\\ Barcelona, Spain\\Email: isuciu@worldsensing.com}
\and
\IEEEauthorblockN{Xavier Vilajosana}
\IEEEauthorblockA{Open University of Catalonia\\ Worldsensing\\ Barcelona, Spain\\ Email: xvilajosana@uoc.edu}
\and
\IEEEauthorblockN{Ferran Adelantado}
\IEEEauthorblockA{ Open University of Catalonia\\
Barcelona, Spain\\Email: ferranadelantado@uoc.edu
}}


%


\maketitle

\begin{abstract}
Low Power Wide Area Networks (LPWANs) are gaining ground in the IoT landscape and, in particular, for Industrial IoT applications. However, given the strict duty cycle restrictions (e.g. 1\% in SubGHz bands) and the limited power supply of devices, requirements of some applications can not always be met. This paper analyzes the potential of the combination of  packet fragmentation -in the direction of the IETF LPWAN working group- and negative group acknowledgement (NACK) in LoRaWAN networks, a widespread LPWAN technology. Results show that the proposed strategy can lead to significant gains in terms of goodput and energy efficiency under congested situations. 
\end{abstract}


%
\IEEEpeerreviewmaketitle

\section{Introduction}
\IEEEPARstart{L}{PWAN}s are star-topology networks composed of battery-operated devices mostly deployed in harsh environments, where the battery replacement is costly. As these devices are required  to provide 10 years of network lifetime, the data they send to the gateway consists in a few packets/day, most of the times without being acknowledged. This is a way to maintain the nodes more time off and also to satisfy the 1\% duty-cycle restrictions imposed by ETSI for license-free bands \cite{etsi-rule}.

The demands for the industrial IoT technologies incremented from the support of simple monitoring applications with low traffic needs to applications where a high quality of service is required for a massive number of low power connected devices \cite{5GPPP}. In this context, the technology provider's aim is to evolve the communication technology towards more reliable and scalable long range wireless by adopting new access mechanism or using dedicated bands~\cite{ neul, rpma, nbiot}. However, we foresee opportunities to leverage the combination of packet fragmentation and group NACK to improve the network scalability that have not been studied. 

The group NACK combined with packet fragmentation will only acknowledge a packet after all its corresponding fragments have been sent and only if there are fragments that need to be resent.  This brings reliability while reduces the impact of individual fragment acknowledgements in terms of duty cycle and energy consumption. Yet, packet fragmentation opens up new challenges and opportunities to be explored for improving the efficiency of these very restricted networks under congestion situations\cite{adelantado17lorawan}.

Packet fragmentation has been traditionally seen as an adaptation mechanism to divide MAC layer Service Data Units (SDU) into a set of smaller Physical layer Protocol Data Units (PDU) with a dual purpose: i) better adapt to the channel conditions by reducing the length of the PDU in noisy channels, and ii) fit long SDUs into maximum length PDUs. However, the impact of an aggressive packet fragmentation strategy in strict duty cycle and energy constrained networks such as LPWANs has not been analyzed in depth. An aggressive packet fragmentation consists in using packet fragmentation, despite a frame fits into the PDU. This strategy could be a way to take better advantage of the available channels in the network, as the smaller the fragments, the shorter the time on air and the higher the opportunity to transmit without collisions. For multi-channel networks, using packet fragmentation spreads the transmission of a packet over a set of channels in a more homogeneous way, thereby allowing channel hopping by fragment.
 Also, in case of fragment/s loss, there is no need to retransmit the entire packet but, only the lost fragment/s, leading to energy savings.

Despite the potential gains of packet fragmentation, some drawbacks arise. First, packet fragmentation could incur energy and communication overhead due to additional fragment headers \cite{suciu17frag1}; secondly, there is an increase in the number of access attempts in the network.

The aim of this paper is to shed light on the potential gains of packet fragmentation combined with group NACK in duty cycle restricted LPWANs and show in which network conditions this strategy is advisable. From the best of our knowledge, similar studies have not been done in the existing literature. The analysis carried out in the sequel is based on the LoRaWAN networks, one of the most adopted technologies for the industrial IoT applications \cite{adelantado17lorawan} that provides very low data rate, ranging from 0.3 kbps to 27 kbps.

The reminder of this article is organized as follows: Section \ref{sec:protdesc} provides a description of LoRaWAN and of the agressive fragmentation protocol we implemented. Section \ref{sec:eval} describes the metrics used and the simulation setup. Section IV provides a discussion of the results, while Section V concludes the paper.

\section{LoRaWAN operation and Aggressive Fragmentation Strategy}
\label{sec:protdesc}
\subsection{LoRaWAN operation}
LoRaWAN is characterized by its PHY layer, namely LoRa, which is a proprietary Chirp Spread Spectrum (CSS) modulation scheme developed by Cycleo, and lately acquired by Semtech, with 125 kHz, 250 kHz or 500 kHz bandwidth and a variable Spreading Factor (SF) with values from 7 to 12 \cite{SemtechWhatIsLoRa}. For a given packet size, both the bandwidth and the SF determine the time required to transmit the packet, also known as Time on Air ($T_{oA}$). Regarding the Medium Access Control (MAC) of LoRaWAN, it is based on the pure ALOHA random access combined with a duty cycle per channel, which for instance in Europe is set to 1\% for the 868MHz ISM band \cite{etsi-rule}. That is, upon the generation of a packet, the node transmits the packet only if there is a channel available for transmission. Yet, the availability of a channel is defined based on its duty cycle. Specifically, in a channel with a duty cycle $DC$ and for a packet with Time on Air $T_{oA}$, the channel only becomes available for the node after an off period, namely $T_{off}$, equal to:
\begin{equation}
T_{off} \text{ [sec]}=T_{oA} \times \frac{100-DC}{DC}
\end{equation}

When more than one channel is available, the node randomly selects the channel to transmit the packet.
\subsection{Aggressive Fragmentation Strategy}

The aggressive fragmentation strategy consists in using packet fragmentation even if the frame fits the PDU, in order to make use of the advantages that come out of using smaller data size \cite{suciu17frag1}. For enhanced network performance, we propose a group-NACK scheme, allowing for fragment retransmissions.

According to LoRaWAN specification \cite{lorawan_spec}, the payload of a packet needs to be sent together with a frame header and a MAC header.
The MAC header (1B) contains 3 bits identifying the message type, 2 bits for the major version of the frame format and  3 bits that are reserved for future use. The frame header (7-21B) uses 4B for the device address, 1B for frame control, 2B as frame counter and up to 15B as frame options.

When the aggressive fragmentation strategy is used, the payload of the generated packet is divided into a set of equal size fragments. To each fragment, a 9B header is added, accounting for the MAC and frame headers. 

Throughout this paper, in order to determine in which network conditions the aggressive fragmentation strategy is advisable, the following transmissions strategies will be analyzed:
\begin{itemize}
\item Aloha: represents the baseline protocol; the data packets are sent unfragmented and only if the channel is available for transmission, otherwise the packets are discarded. 
\item Buffered Aloha: the data packets are buffered until a channel becomes available and then sent consecutively, unfragmented and subject to the duty cycle restrictions of the network .
\item Buffered Aloha with fragmentation: the data packets are fragmented and buffered until the channel becomes available for transmission; the fragments are sent consecutively and subject to the duty cycle restrictions of the network. If after all the fragments of a packet have been sent, at least one of the fragments is lost, the whole packet is dropped by the gateway.
\item Buffered Aloha with fragmentation and retransmissions: in this case, after all the fragments of a packet have been sent, the node waits for a NACK. The NACK indicates which fragments have not been received. In case a NACK is received, it will proceed with resending the missing fragments, following the same protocol and respecting the duty cycle restrictions of the network. If even after the corresponding retransmission sessions for that packet at least one of the fragments is still lost or corrupted, the whole packet is dropped by the gateway.
\end{itemize}

The NACK should contain a MAC and frame header in order for the node to identify if the message is meant for it. There can also be a payload attached to it, of variable size. Our choice was to map the fragments status in the NACK on a 0-1 basis, in function of the sequence number of the fragments: 0-if the fragment was not received and needs retransmission and 1 if it was correctly received at the gateway. This strategy needs the GW to be aware of the number of fragments that the nodes in the network use and that all the nodes use the same number of fragments/packet. Also, the retransmission of a fragment is made using the same sequence number as it had when it was first sent, so that this mapping can be correctly updated.

\begin{figure}[htbp]
 \centerline{\includegraphics[width=79mm]{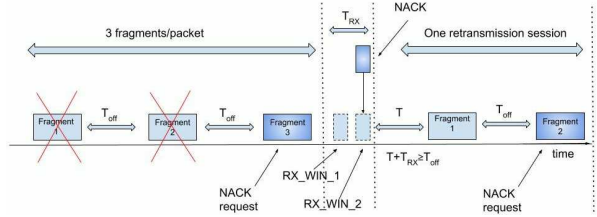}}
\caption{Sending a packet using 3 fragments: the last fragment is the one requesting a NACK. If a NACK is sent by the gateway, it is sent during the first or second receive window opened by the sensor node. The two failed fragments will be sent as soon as possible, after the mandatory $T_{off}$ expires. The last fragment sent can request again for a NACK, if more retransmission sessions per packet are wanted.}
\label{NACK_expl}
\end{figure}

For the retransmission protocol, we are proposing a scheme in which the last fragment of a packet will be the one triggering the NACK request. The NACK can be received in one of the two reception windows that will be opened by the sensor node  after the UL data is sent, as described by LoRaWAN \cite{lorawan_spec}. In case this last fragment is lost, there will be no NACK and the node will continue its activity by sending other packets. If a NACK is received, the node will start sending the fragments that are marked as lost. All these lost fragments that are being resent correspond to one 'retransmission session', as shown in \fig{NACK_expl} for the case of a network configured to use 3 fragments/packet.

We chose to implement the retransmission scheme in this way because these networks are restricted by the duty cycle and by the energy consumption: choosing to ACK each fragment or packet and retransmitting until the ACK is received is too expensive in both duty cycle and energy consumption \cite{raza17bidirectional}. This scheme ensures that the nodes will resend only if they are explicitly told so.

\section{Performance Evaluation}
\label{sec:eval}

\begin{figure}[htbp]
 \centerline{\includegraphics[width=80mm]{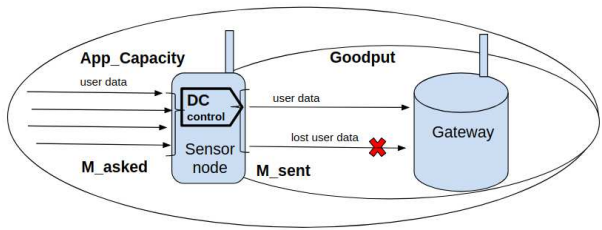}}
\caption{Behavior of a sensor node operating in a duty cycle restricted network: data can only be sent to the GW when the DC allows it.}
\label{DC_bottleneck}
\end{figure}

In duty cycle restricted networks, after a node sends data, it has to stay silent for the mandatory $T_{off}$ corresponding to that data, as defined in Section \ref{sec:protdesc}. This means that if the IoT application running on that node asks for more data during $T_{off}$, the node will drop that data (Aloha) or will buffer it for until it is allowed to send again (Buffered Aloha). This is managed in \fig{DC_bottleneck} by the 'DC control' module.

The first part of this section presents the metrics that we used for the performance evaluation of the transmission strategies presented in Section \ref{sec:protdesc}, while the second part discusses details of the network simulations we developed using NS3. 

\subsection{Performance metrics}

The performance of LPWANs is mainly evaluated in terms of goodput and energy consumption. These, as well as other metrics used in this paper are described in the following.


\textit{Goodput}: It is defined as the percentage of packets correctly received by the gateway, with respect to the amount of data sent in the network. It is expressed as
\begin{equation}
Goodput [\%]=\frac{M_{c}}{M_{sent}} \times 100
\end{equation}
\noindent where $M_{sent}$ corresponds to the number of data packets sent in the network and $M_c$ to the number of packets correctly received by the gateway. In case of fragmentation, the packet is only received correctly if all  its fragments have been correctly received. $M_{sent}$ does not account for packet retransmissions. 

\textit{Application Capacity}: It is defined as the percentage of packets correctly received by the gateway, with respect to the amount of data asked by the application. This metric allows us to identify the region starting with which packet fragmentation brings a gain to the network performance, despite the headers overhead. It is expressed as
\begin{equation}
App\text{ }Capacity [\%]=\frac{M_{c}}{M_{asked}} \times 100
\end{equation}
\noindent where $M_{asked}$ corresponds to the number of data readings asked by the IoT application. This data may not all be sent in the air interface because of the DC restrictions of the network (\fig{DC_bottleneck}). $M_c$ is the same parameter as defined for Goodput. 


\textit{Energy Efficiency}: it is defined as the total energy consumption of the network divided by the number of successful packets delivered by the sensor nodes to the gateway:
\begin{equation}
\label{eq:EnEff}
Energy\text{ } Efficiency \text{ [J/packet]}=\frac{E}{M_{c}}
\end{equation}
\noindent where $E$ is the energy consumption of the network and $M_{c}$ represents the number of correctly received packets at the gateway, as defined for Goodput. The energy consumption of the network accounts for the processes of sending data (packets, fragments, headers) and for processing the NACKs, if it is the case.

\textit{Header overhead}: This overhead is caused by the need to transmit an additional header for each fragment, as described in Section \ref{sec:protdesc}. In order to assess this impact, we define the fragmentation header overhead as the percentage of extra energy devoted to transmit a packet in a certain amount of fragments compared to the energy required to transmit the packet in one piece. Therefore,
\begin{equation}
\label{eq:HeaderOH}
Header \text{ }Overhead \text{ [\%]}= \Bigg(\frac{n_f * E_{f}}{E_{packet}} -1 \Bigg)\times 100
\end{equation}
\noindent where $n_f$ is the number of fragments required for sending a packet, $E_f$ is the energy required to transmit one fragment of the respective size and $E_{packet}$ is the energy required to transmit the packet unfragmented. $E_f$ and $E_{packet}$ are proportional to their corresponding transmission duration.


\subsection{Simulation Setup}

The simulations have been developed using the NS3 network simulator. We evaluated our approach with network sizes ranging from 1 to 50 sensor nodes for a single gateway and fixed coverage area. 

In order to assess the performance of a \textit{dense} network, we chose having all the nodes operating in a single channel and with the same SF: the network operates in a channel of 125 kHz bandwidth in the 868 MHz ISM band and all the nodes transmit with SF=7. The NS3 simulator evaluates the network performance by taking into account not only the packets destroyed by collisions but also the ones destroyed by interference or having a power below the sensitivity threshold of the gateway. 


The IoT application running on each node will ask for a fixed amount of data, $M_{asked}$, independent of the transmission strategy. Because of the DC restrictions of the network, only $M_{sent}$ out of $M_{asked}$ will be delivered to the gateway (as shown in \fig{DC_bottleneck}).

The data packets have a fixed payload of 200 B, close to the maximum size that LoRaWAN can send using SF7 \cite{lorawan_spec}. If considering other values for the SF, the payload should be modified accordingly so as to be close to the maximum allowed value \cite{lorawan_spec}. In this way, the protocols described in Section \ref{sec:protdesc} can be evaluated: Aloha, Buffered Aloha, Buffered Aloha with fragmentation and Buffered Aloha with fragmentation and retransmissions. 

Whenever the fragmentation option is used, each data packet will be split into 2 to 5 fragments, but all the sensor nodes in the network will use the same number of fragments/packet. The gateway keeps track of the arrived fragments from the sensor nodes and will be able to provide them with a Group-NACK per packet. After the maximum number of retransmission sessions is completed, the gateway will discard the packets that still have missing fragments. A retransmission session means sending all the fragments that a NACK marked as lost or damaged (see \fig{NACK_expl}).

\section{Results}

In the following, the metrics defined in Section  \ref{sec:eval} are analyzed in order to determine if and when the aggressive fragmentation  strategy is advisable for the case of duty cycle restricted LPWANs.
\subsection{Network Goodput}

\begin{figure}[htbp]
  \centerline{\includegraphics[width=80mm]{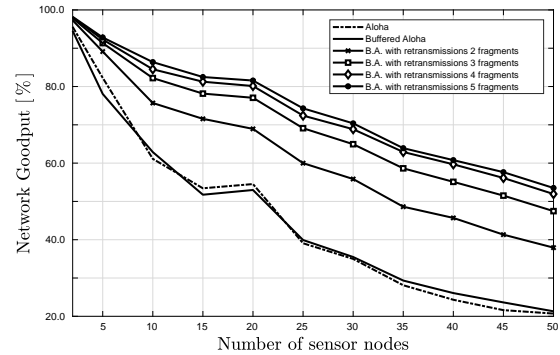}}
\caption{The variation of the network goodput with an increasing number of sensor nodes in the network. Transmission Strategies: Aloha, Buffered Aloha and Buffered Aloha with fragmentation and one retransmission session per packet (2, 3, 4 and 5 fragments/packet).}
\label{good2}
\end{figure}

The network goodput (\fig{good2}) starts with a value of 100\% for any transmission strategy when there is only one device in the network. This value decreases as the number of devices (and collisions) in the network increases. Aloha and Buffered Aloha (B.A) will deliver almost the same goodput performance, as they only differ in timing.

\begin{figure}[htbp]
  \centerline{\includegraphics[width=80mm]{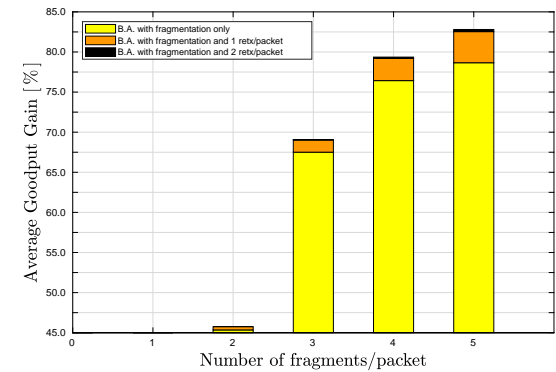}}
\caption{The average gains obtained in network goodput with respect to only using Buffer Aloha transmission. Transmission strategies: fragmentation in 2 to 5 fragments/packet, fragmentation and 1 retransmission session/packet and fragmentation and 2 retransmission sessions/packet.}
\label{good2_gains}
\end{figure}

When using B.A with fragmentation and retransmissions, the variation of the network goodput with increasing number of devices becomes smoother. Also, the higher the number of fragments/packet, the higher the increase in goodput, as more correct packets are delivered correctly to the gateway. This happens because having smaller data packets reduces the probability of collisions and increases the probability of receiving NACKs (\fig{good2_gains}).

In \fig{good2} we couldn't show both the case of B.A with fragmentation only and B.A with fragmentation and retransmissions, as the scale didn't allow us. So, \fig{good2_gains} shows the gains in goodput that are obtained compared to Buffered Aloha, when using the B.A with fragmentation only policy. On the same figure, there are plotted the extra gains obtained when upgrading to B.A with fragmentation and one retransmission session/packet, followed by the gains brought by using 2 retransmission sessions/packet. 

As we can see, using 5 fragments/packet and 1 retransmission session/packet brings in average an additional 4\% gain to using B.A with fragmentation only. Moreover, using 2 retransmission sessions/packet brings additional gains that are smaller than 0.5\%  and happen only for configurations of more than 3 fragments/packet. This is why the remaining of the paper will not treat the case of  using 2 retransmission sessions/packet.

\subsection{Application Capacity}

\begin{figure}[htbp]
  \centerline{\includegraphics[width=80mm]{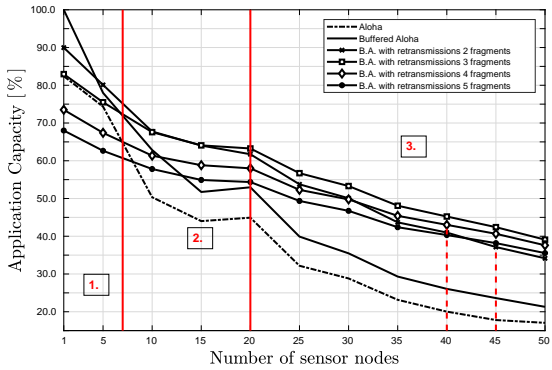}}
\caption{The variation of the application capacity with increasing number of devices in the network. Transmission Strategies: Aloha, Buffered Aloha and Buffered Aloha with fragmentation and one retransmission session per packet (2, 3, 4 and 5 fragments/packet).}
\label{app_cap}
\end{figure}

\fig{app_cap} shows the variation of the application capacity with an increasing number of devices operating in the same channel and using the same SF. This metric helps us identify the network conditions in which the packet fragmentation strategy becomes helpful.

For a small network load (region marked as '1' in \fig{app_cap}), data can be sent using full packet size (in our case, 200B). This strategy provides the best results because the probability of collision is low, so using fragmentation would add overheads that are not necessary. Aloha provides worse results than Buffered Aloha, as it is wasting the time resource of the network, directly affecting the application capacity of sending user data. 

The second region of the plot shows that sending data using 2 fragments/packet is the strategy leading to the best obtainable results. With an increasing number of devices in the network and increased number of collisions, the third region is the one where sending 3 fragments/packet exceeds the other transmission strategies. The two dashed lines in the plot mark the regions where sending data in 4 fragments/packet and 5 fragments/packet, respectively, overtake the performance provided by using 2 fragments/packet. Still, they cannot exceed the application capacity corresponding to 3 fragments/packet. This happens because the lowering in probability of collision that they cause it is not important enough so as to make up for the fact that their extra $T_{oA}$ directly affects the application capacity.

Going back a step, \fig{good2} showed us that the smaller the data size the better the network goodput obtained. Now, \fig{app_cap} shows us that depending on the region in which the network operates, there is a trade-off needed in the number of fragments/packet to be used, so that fragmentation doesn't have a negative impact on the application capacity.


\subsection{Energy efficiency}
\begin{figure}[htbp]
 \centerline{\includegraphics[width=80mm]{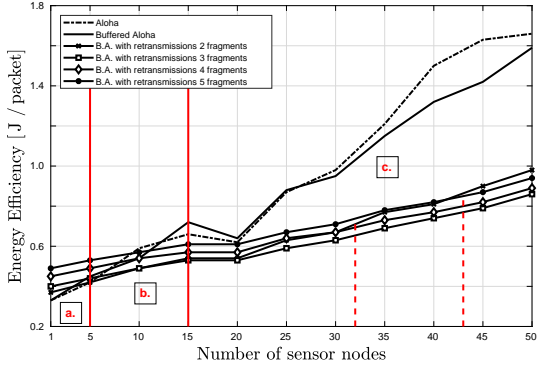}}
\caption{The energy efficiency of Aloha, Buffered Aloha and Buffered Aloha with fragmentation and one retransmission session per packet (2, 3, 4 and 5 fragments/packet)}
\label{en2}
\end{figure}

The energy efficiency of the network (\fig{en2}) follows a similar trend with the application capacity, but it is strongly dependent on the network goodput (amount of data sent, amount of data correctly received by the gateway). The region marked with 'a' corresponds to Aloha as being the most energy efficient protocol. This happens because Aloha sends less data than Buffered Aloha. Using packet fragmentation and retransmissions in this region is not recommended, as this would imply extra energy consumption for providing a similar network goodput.

The 'b' region shows a number of 2 fragments/packet as being the most energy efficient strategy, very close to the performance that using 3 fragments/packet provides. This happens because the extra energy consumption of using 3 fragments/packet is compensated by the goodput improvement that this strategy brings.  

For the networks operating in the 'c' region, using 3 fragments/packet is a good trade-off between the network energy consumption and the obtained goodput performance. The two dashed lines have the same significance as for \fig{app_cap}.

We see that Aloha and Buffered Aloha have the worst energy efficiency for dense networks. Using 4 or 5 fragments/packet would provide a better network goodput than using a lower number of fragments/packet, but a price needs to be paid in terms of energy-efficiency.


\subsection{Header overhead}
In TABLE I, the overhead that packet fragmentation brings in terms of $T_{oA}$ and implicitly, energy consumption is computed. In the middle column, 9B headers are assumed for each fragment, while in the left column we consider 1B headers. 
If a way to shrink the 9B MAC and frame header into a 1B fragmentation header (in the direction of the IETF LPWAN working group) is found, the energy efficiency of the network would be improved.
\begin{table}[h]	
\caption{\label{header_ovhd}Header overhead  for multiple fragmentation options}
	\centering
	\begin{tabular}{p{1.8cm}||p{2.4cm} |p{2.4cm}}
		\hline Fragments/packet &  Header impact/packet [9B] & Header impact/packet [1B] [\%]\\
        \hline
		\hline 2 &  8.93 & 5.71\\ 
		\hline 3 &  19 & 12.61\\ 
        \hline 4 &  26.8 & 17.14\\ 
		\hline 5 &  35.71 & 22.86\\ 
		\hline
	\end{tabular}
\end{table}

\section{Conclusion}
In this paper we proposed a transmission strategy that combines packet fragmentation with group NACK  in duty cycle restricted LPWANs. Packet fragmentation is used despite the packet fits the frame, so as to reduce the probability of collisions while the number of users in the network increases. The group NACK is requested by the last fragment of a packet and accounts for all the fragments of that data packet.  This strategy is shown to provide increased network goodput and energy efficiency for dense networks. The retransmission policy is more efficient for smaller fragment sizes, where the probability of successful NACK request is higher. 

We provided insights so as to show what transmission strategies are advisable as function of the network size. We showed that for small networks, it is better not to use packet fragmentation, but to use Aloha or Buffered Aloha, which provide similar goodput at increased application capacity and energy efficiency. This is true also for IoT applications that only need to send packets of very small payload, below the size of any fragment considered in this work.  

As the network size increases, the aggressive fragmentation strategy provides better network performance. The number of fragments/packet to be used could be dynamically adapted so as to provide the best network performance: goodput, application capacity or energy efficiency. Here, there is a trade-off that needs to be done: smaller fragments provide better goodput but they are less energy efficient and decrease the IoT application capacity. This is mainly because of the fragmentation headers that represent a high overhead in terms of extra time on air and energy consumption. The gateway could control the number of fragments/packet that the nodes use by issuing a MAC command.

The performance of dense industrial duty cycle restricted LPWANs could be further improved if a more collision-resilient acknowledgement scheme is found and if the fragmentation header size is reduced. 


\section*{Acknowledgment}
This project has received funding from the European Union Horizon 2020 research and innovation programme under the Marie Sklodowska-Curie grant agreement No 675891 (SCAVENGE project). This work is also partially supported by the Spanish Ministry of Economy and the ERDF regional development fund under SINERGIA project (TEC2015-71303-R).



%


\bibliographystyle{IEEEtran}
\bibliography{IEEEabrv,PIMRC_conf}

\end{document}